\documentclass[aps,prb,twocolumn,showpacs,superscriptaddress,floatfix]{revtex4-1}

\usepackage{amsmath}
\usepackage{amssymb}
\usepackage{graphicx}
\usepackage{grffile} % to handle dots in graphics filnames
\usepackage{float}
\usepackage{color}
\usepackage{subfigure}

\newcommand{\kk}{\mathbf{k}}
\newcommand{\A}{\mathbf{A}}
\newcommand{\EE}{\mathbf{E}}

\newcommand{\trel}{t_\mathrm{rel}}
\newcommand{\tave}{t_\mathrm{ave}}

\newcommand{\tmin}{t_\mathrm{min}}

\begin{document}

\author{A.~F.~Kemper}
\email{afkemper@lbl.gov}
\affiliation{Lawrence Berkeley National Laboratory, 1 Cyclotron Road, Berkeley, CA 94720, USA}

\author{M.~A.~Sentef}
\affiliation{Stanford Institute for Materials and Energy Sciences (SIMES),
SLAC National Accelerator Laboratory, Menlo Park, CA 94025, USA}

\author{B.~Moritz}
\affiliation{Stanford Institute for Materials and Energy Sciences (SIMES),
SLAC National Accelerator Laboratory, Menlo Park, CA 94025, USA}
\affiliation{Department of Physics and Astrophysics, University of North Dakota, Grand Forks, ND 58202, USA}

\author{J.~K.~Freericks}
\affiliation{Department of Physics, Georgetown University, Washington, DC 20057, USA}

\author{T.~P.~Devereaux}
\affiliation{Stanford Institute for Materials and Energy Sciences (SIMES),
SLAC National Accelerator Laboratory, Menlo Park, CA 94025, USA}
\affiliation{Geballe Laboratory for Advanced Materials, Stanford University, Stanford, CA 94305}

\title{Effect of dynamical spectral weight redistribution on effective interactions in time-resolved spectroscopy}

\begin{abstract}
The redistribution of electrons in an ultrafast pump-probe experiment causes significant changes to the spectral distribution of the retarded interaction between electrons and bosonic modes.  We study the influence of these changes on pump-probe photoemission spectroscopy for a model electron-phonon coupled system using the nonequilibrium Keldysh formalism.  We show that spectral rearrangement due to the driving field preserves an overall sum rule for the electronic self-energy, but modifies the effective electron-phonon scattering as a function of energy.  Experimentally, this pump-modified scattering can be tracked by analyzing the fluence or excitation energy dependence of population decay rates and transient changes in dispersion kinks.
\end{abstract}

\maketitle

\section{Introduction}
Imaging electron dynamics on femtosecond time scales, on which many-body effects occur, has become a reality given technical advances of a number of pump-probe techniques.
The aim of pump-probe studies is to learn about systems with coupled electronic, lattice, charge, and spin
degrees of freedom by driving them out of their equilibrium state and studying their
relaxation dynamics.  
This has been applied to a large variety of materials with a variety of probes, including
quantum dots\cite{p_kirchmann_10},
correlated oxides\cite{a_cavalleri_01,c_kubler_07,e_abreu_12,coslovich_13}
topological materials\cite{niesner_12,j_sobota_13,y_wang_13}
charge- and spin-density wave materials\cite{Schmitt2008,l_stojchevska_10,s_hellmann_12},
and unconventional superconductors\cite{j_graf_11,l_rettig_12,w_zhang_13,j_rameau_14}.

The pump is a strong electric field that causes a number of changes in the
system.  On a basic level, it can excite states through either dipole absorption or by driving
electrons within their respective bands, redistributing the charge density.  In more complex
systems, it can lead to changes in the orbital/spin character of the underlying quasiparticles,
or break pairs in systems with emergent order such as density waves and superconductivity.
With sufficiently strong fields, even new states of matter can be induced that do not exist in equilibrium
in certain systems such as graphene and topological insulating materials.\cite{y_wang_13}
The pump will clearly modifies the interactions between the constituents.

In this work, we consider the effects of pumping a simple system of electrons coupled to lattice
vibrations (phonons) to answer the question of how their mutual interaction is changed by the pump.
We show that the pump-induced redistribution of electrons leads to \emph{apparent} modifications of the electron-boson coupling even though a sum rule guarantees that the total interaction strength is preserved for all times. 
A common method for analyzing time-resolved phenomena is to refer back to equilibrium models
with modified or time-dependent parameters; this approach can be appropriate at the latest stages
of thermalization\cite{m_sentef_13}.
The modification of electron-boson coupling found here cannot
be captured in such a picture for all times, and needs a more careful treatment, which should provide a framework
for analysis in experiments.

Here, we focus on changes induced by driving-field modification of the electronic distribution.  As the electron
phonon interaction is in part determined by the distribution of electrons, any rearrangement of spectral weight
will lead to a modified effective interaction.  
These effects are illustrated by simulating tr-ARPES spectra and analyzing the many-body self-energy and
signatures of the electron-phonon interaction in the extracted band dispersion and quasiparticle widths or lifetimes.
The weakening
of spectral kinks has been observed experimentally in a cuprate material, where it was attributed
to light-induced changes in the interactions.\cite{j_rameau_14}

We find at short times that while the interactions are modified, the system possesses a sum rule for the energy integrated self-energy which characterizes the overall electron-phonon coupling in the system.  Although the shift leads to a weakening of some features associated with the coupling in equilibrium (i.e. bosonic ``kinks'' and line widths), the electrons do not decouple from the phonons.  Rather, the interaction is modified by redistributed spectral weight, and the apparent loss of interaction strength at low frequency
is compensated for elsewhere.

\section{Model and methods}

We solve the time-dependent equations of motion for the Holstein model\cite{Holstein}
\begin{align}
\mathcal H= \sum_\kk \epsilon(\kk) c^\dagger_\kk c_\kk + \Omega  \sum_i b_i^\dagger b_i - 
		g \sum_i c_i^\dagger c_i \left( b_i + b_i^\dagger \right)  \nonumber
\end{align}
where the individual terms represent the kinetic energy of electrons with a dispersion $\epsilon(\kk)$, the energy of phonons with a frequency $\Omega$, and a coupling between them whose strength is given by $g$. 
Here, $c^\dagger_\alpha (c_\alpha)$ are the standard creation (annihilation) operators for an electron in state $\alpha$; similarly, $b^\dagger_\alpha (b_\alpha)$ creates (annihilates) a boson in state $\alpha$.
We use a square lattice dispersion with nearest neighbor hopping ($V_{nn})$,
\begin{align}
\epsilon(\kk) = &-2 V_{nn} \left[ \cos(k_x) + \cos(k_y)\right] -\mu
\end{align}
where $\mu$ is the chemical potential.  We have used the convention that $\hbar=c=e=1$.
%, which makes inverse energy the unit of time.  

The interactions are treated within the self-consistent
Born approximation, and the equations of motion (detailed in Appendix~\ref{sec:app_model}) are solved
using the non-equilibrium Keldysh formalism\cite{stefanucci_book}.
The driving field is included through the Peierls' substitution\cite{r_peierls_33} $\kk(t)= \kk - \A(t)$, where $\A(t)$ is the vector potential assumed to be spatially uniform.  We work in the Hamiltonian gauge with no scalar potential $\Phi$, so the electric field is obtained via $\EE(t) = -\partial_t \A(t)$, and model the propagating pump pulse using a single central frequency and a Gaussian envelope.  
The pump and probe field profiles are taken to be
relevant to current experiments, with achievable frequencies and durations.
A further rule of thumb is that the units for electric field (here in eV)  are related to the true electric field by electron charge $e$ and the lattice constant $a_0: \EE\rightarrow \EE/(e a_0)$ (since $\A$ is in units of $a_0^{-1}$).  

The equilibrium coupling strength is characterized by the dimensionless parameter
$
\lambda \equiv -\partial \mathrm{Re} \Sigma^R(\omega) / \partial \omega |_{\omega\rightarrow 0}
$
which depends on the bare interaction vertex $g$, electronic density of states, and temperature.
The system parameters are chosen to represent a generic electron-phonon system, where the coupling is sufficiently weak that the system
can be described at the Migdal level;
the dispersionless phonon frequency $\Omega$ is small compared to the bandwidth $W$ $(\Omega\ll W)$.

The parameter choices are motivated by a combination of physical
relevance and computational feasibility.  To clearly separate the energy scale of the phonon kink from
the Fermi level, the phonon frequency used is $\Omega=0.2$ eV, where the coupling strength is chosen
sufficiently low to remain within the Migdal limit ($\lambda \lesssim 1$).  
In materials, phonon frequencies tend to be lower \textemdash the results demonstrated below
occur for lower frequencies as well.
We further consider a half-filled system ($\mu=0$ eV) with a hopping strength $V_{nn}=0.25$ eV.
Below, we will use two sets of parameters for the coupling strength and pump, 
chosen to emphasize certain aspects.
In the first case, where we focus on the kink dynamics, we excite roughly 10\% of the electrons in the
band to above the Fermi level.  
In the second case, where we study the fluence dependence of the decay rates, a maximum of 5\% of the
electrons in the band are excited (for the largest fluence).

The first effect is the considerable reduction in the electron-phonon kink strength.
To emphasize the kink and the changes upon pumping, we have used
a system with moderate coupling strength, where
set $g = \sqrt{0.12}$ eV  and $T\approx83$K ($\lambda\approx0.7$), and a strong driving field.  The
driving field used here was
 $A_\mathrm{max}=0.612 /a_0$ at a frequency $\omega=0.1$ eV, corresponding to a maximum electric field of 61.2 mV$/a_0$ (where $a_0$ is the lattice constant).  The pump duration was $32.9$ fs.
The second effect is the modification of the decay rates due to the pump.  Here, the coupling strength
above is too large to reliably extract exponential decay rates since the system 
rapidly decays at large energies. Due to the low temperature, the decay rate is conversely
too low at low energies.
%evolves into a
%state where the only remaining changes are inside the phonon window 
%even before the pump pulse is fully ``off''.  
Thus, we decreased the effective high-energy decay rates by changing
the coupling to $g=\sqrt{0.016}$ eV  and increased the effective low-energy rates by increasing
the system temperature to $T=290$K ($\lambda \approx 0.1$).
Additionally, due to computational constraints on the contour length, the pulse was compressed by
increasing the driving frequency to $\omega=0.5$ eV, and the width to $6.58$ fs.

\begin{figure*}[thb]	
	\includegraphics[width=0.8\textwidth]{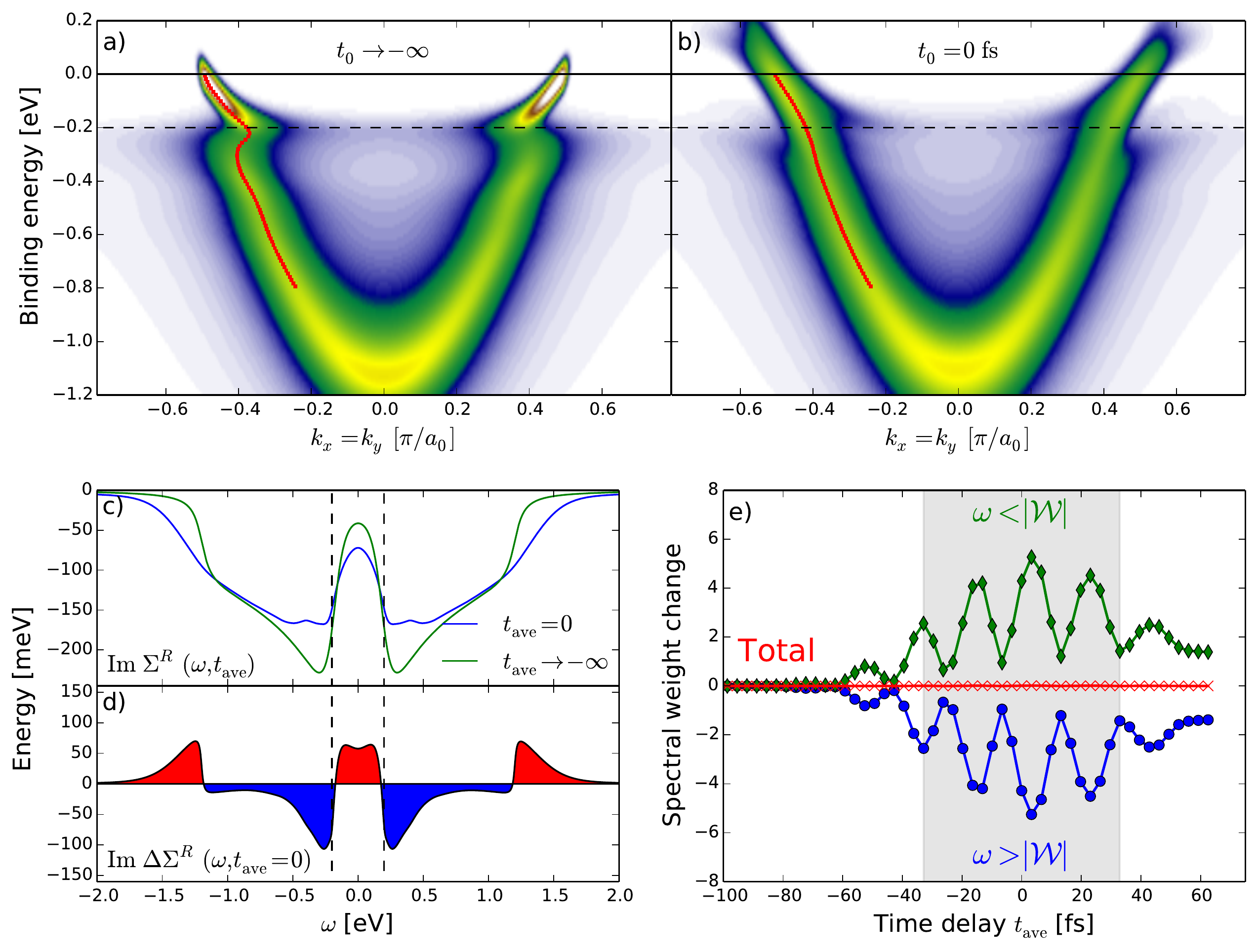}
	\caption{tr-ARPES spectra along the $k_x=k_y$ direction (a) in equilibrium and (b) at 0 time delay.  Red points indicate peaks in the MDC curves as determined from Lorentzian fits, and the black dashed line indicates the phonon window $\mathcal{W}$.  (c) Imaginary part of the Wigner self-energy $\Sigma^R(\omega,\tave)$ corresponding to panels (a) and (b).
	(d) Spectral weight change in $\mathrm{Im}\ \Sigma^R(\omega,\tave)$ at $\tave=0$.  (e) Time evolution of the shift in spectral weight of $\mathrm{Im}\ \Sigma^R(\omega,\tave)$ within (green) and outside (blue)  $\mathcal{W}$.  The red line indicates the total change in spectral weight, which is zero due to the sum rule.  The grey area shows the region where pump effects are present (see text).
	}
	\label{fig:fig1}
\end{figure*}

\section{Results}
We begin by studying a strongly coupled system at low temperatures, excited by a strong driving field. Figures~\ref{fig:fig1}(a) and (b) show the tr-ARPES spectra $I(\kk,\omega,t_0)$ at two time delays, prior to and during the pump pulse, respectively.  Overlaid on the spectra are red dots indicating peaks in Lorentzian fits to various momentum distribution curves (MDCs) of the data taken at constant binding energy.
The peak in the fit is an indication of the interacting dispersion.
In equilibrium ($t_0\rightarrow -\infty$), the tr-ARPES spectrum shows the characteristics of a strongly coupled Holstein phonon \textemdash\ well-defined spectral peaks at energies within the ``phonon window'' ($\mathcal{W} = \omega \in [-\Omega,\Omega]$) where the linewidth is small, and a strong kink at $\Omega$.  
At zero time delay ($t_0=0$ fs), when there is maximum overlap between the pump and probe, the kink that occurs at the phonon frequency flattens out, indicating an \textit{apparent} change in the 
effective electron-phonon interaction due to the pump.  

With the decrease of the kink, it would appear that 
the underlying electron-phonon interactions have weakened.
To investigate whether this is the case, 
we calculate the retarded self-energy $\Sigma^R(t,t')$ and perform a relative-time Fourier transform (or Wigner transform) to obtain the Wigner self-energy $\Sigma^R(\omega,\tave)$ (see Appendix~\ref{sec:app_model} for details).  Fig.~\ref{fig:fig1}(c) shows $\mathrm{Im}\ \Sigma^R(\omega,\tave)$ in equilibrium and at $\tave=0$ ($\mathrm{Re}\ \Sigma^R(\omega,\tave)$ will be discussed later).  In equilibrium, $\mathrm{Im}\ \Sigma^R(\omega)$ has a region where it is relatively small due to kinematic constraints, i.e. the phonon window.  This phonon window was shown to be responsible for slow decay within some range of the Fermi level in tr-ARPES experiments\cite{m_sentef_13}.  During the pump, electronic spectral weight is redistributed leading to changes in $\mathrm{Im}\ \Sigma^R(\omega,\tave)$. 
Fig.~\ref{fig:fig1}(d) illustrates these changes as differences between the result at $\tave=0$ and in equilibrium.  These changes are positive inside $\mathcal{W}$, as well as beyond the band edges, and negative elsewhere.  However,
the \textit{total} interaction strength is unchanged, which can be seen from Fig.~\ref{fig:fig1}(e), where we plot
the integrated changes both inside and outside $\mathcal{W}$ versus time delay.
The resulting curves show exactly canceling increases and decreases inside and outside the phonon window,
indicating that the total interaction strength remains constant during (and after) the pump.
The oscillations are due to the field profile and the region shown in gray indicates times where the Wigner transform $\tave$ is within one standard deviation of the peak field, i.e. the field is ``on''.

\begin{figure}[ht]	
	\includegraphics[width=\columnwidth]{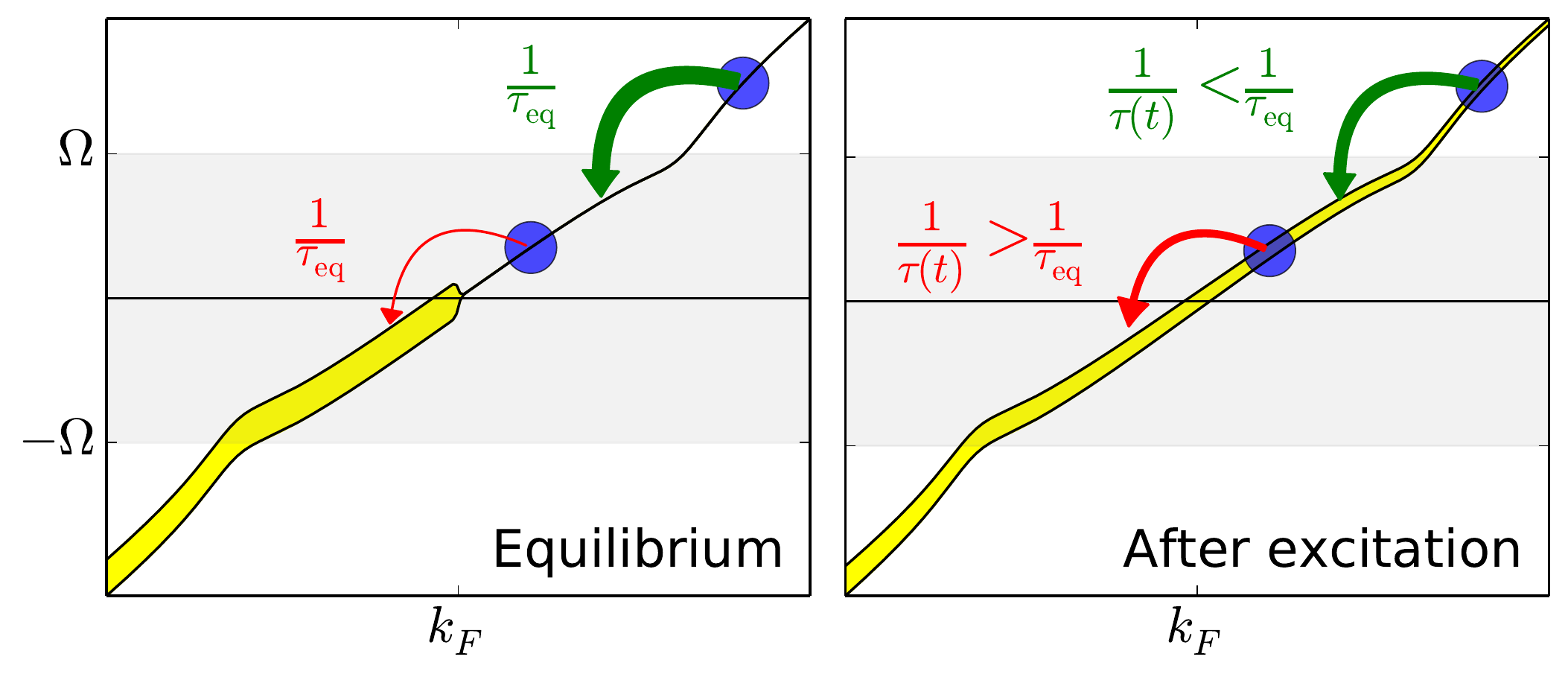}
	\caption{Illustration of the effects on the scattering rate due to spectral weight redistribution.  The thickness
	of the yellow lines indicates the electronic occupation of the band. In equilibrium and at low temperatures, scattering from inside the phonon window $\mathcal{W}$ is suppressed due to a lack of phase space for the final state given the high occupation below the Fermi level, with the opposite behavior for scattering outside the window $\mathcal{W}$.  After excitation and spectral weight rearrangement, these phase space considerations are modified with an increase (decrease) of the scattering rate inside (outside) $\mathcal{W}$.}
	\label{fig:cartoon}
\end{figure}

The constant integrated interaction strength is due to a sum rule for the self-energy,
which can be obtained analytically.
We evaluate the zeroth moment of the
self-energy, which is the equal-time self-energy $\Sigma^R(t,t)$
\begin{align}
\Sigma^R(t,t'=t) = -ig^2\big[2n_B(\Omega/T)+1\big],
\end{align}
where $n_B(x)$ is the Bose function
[the derivation is shown Appendix~\ref{sec:app_sumrule}].
% $n_B(x) = \left[ e^x-1\right]^{-1}$. 
Hence the self-energy, and by extension the total transient effective interaction, obeys a sum rule.\cite{j_freericks_14}  This can be viewed as an alternative measure of the electron-phonon
coupling strength, which is relevant for non-equilibrium physics.

It is important to note that the changes in the tr-ARPES spectra are not caused
by changing the electron-phonon coupling itself, 
known as a quantum ``quench'' where one of the parameters of the system is changed \textit{ad hoc}.  
Instead,  
our self-consistent evaluation of the equations of motion captures the redistribution of spectral weight by the pump and its effects on the transient effective electron-phonon interactions. 
As the spectral weight rearrangement is controlled to a large degree by the pump strength, the signatures of the interaction in the time domain \textemdash the kink and decay rates \textemdash depend strongly on the pump fluence or excitation density.  As the pump fluence increases, spectral weight redistribution increases concomitantly and a time- and fluence-dependent effective electron-phonon interaction emerges.

We can understand changes in the effective electron-phonon interactions by considering the scattering processes in equilibrium and after excitation.  The self-energy in equilibrium is related to the scattering rate via $\mathrm{Im}\ \Sigma^R(\omega)=\left[2\tau(\omega)\right]^{-1}$; out of equilibrium both sides of the equation acquire a non-trivial time dependence and the equation becomes a proportionality due to complications of the Wigner transform
(see Appendix~\ref{sec:app_model} for details).
%\footnote{Due to the difference in the procedures used to obtain either the self-energy or extract the decay rate,
%the full equality involves convolutions over resolution functions as well as relative time Fourier transforms.
%Further details are discussed in the SI}.

%
Fig.~\ref{fig:cartoon} illustrates the scattering processes that occur in phonon emission for both cases.  In equilibrium, the scattering rate of a single excited particle is principally determined by the amount of phase space available for scattering.  A particle that is excited within $\mathcal{W}$ of the Fermi level cannot emit a phonon to scatter because the final states are fully occupied; similarly, a particle that is excited above $\mathcal{W}$ scatters more easily.  It was shown previously that these scattering rates can be quantitatively connected to the equilibrium self-energy in the limit of weak fluence or excitation density.\cite{m_sentef_13}  However, 
when the fluence is increased 
a particle within $\mathcal{W}$ has available phase space to scatter into, leading to an increased scattering rate compared to equilibrium based solely on this redistribution of electronic spectral weight.  On the other hand, particles outside $\mathcal{W}$ now have a decreased scattering rate due to the accessible 
final states being partially occupied.  These changes in the scattering phase space and rates are reflected in Fig.~\ref{fig:fig1}(c) and (d), where $\mathrm{Im}\ \Sigma^R(\omega,\tave)$ increases inside the phonon window, and decreases outside.  In addition to scattering via phonon emission, there will be processes that scatter particles into states at higher energy, i.e. phonon absorption.  However, for low excitation densities these processes will not qualitatively affect the simple picture discussed here.\cite{j_sobota_14}

%
% This is here for formatting
%
\begin{figure}[b]	
	\includegraphics[width=\columnwidth]{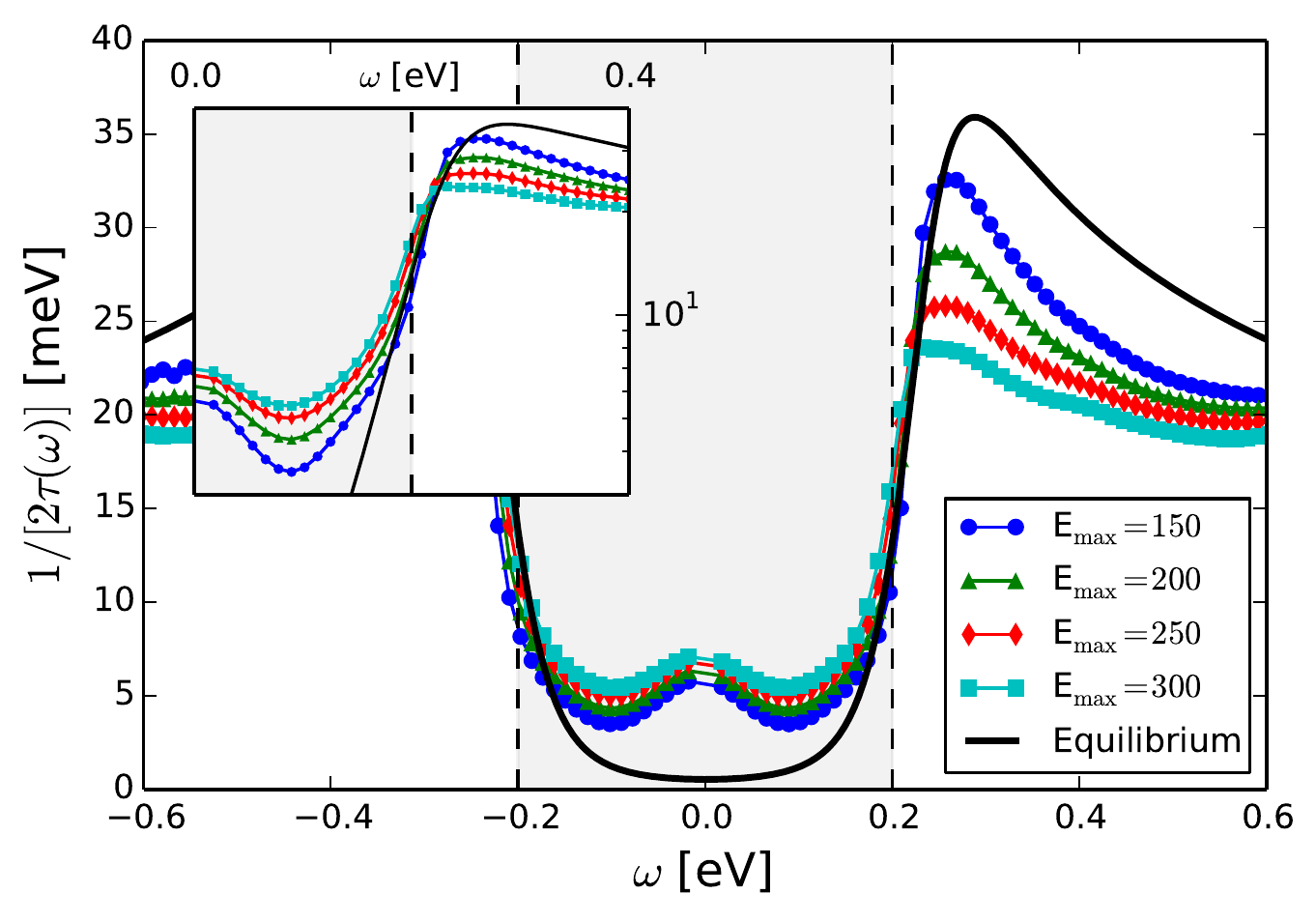}
	\caption{Fluence dependence of the decay rates extracted from $\bar{I}(\omega,t_0)$ near the end of the pump ($t_\mathrm{fit}=46$ fs).  Fields are expressed in units of meV/$a_0$ ($a_0$ is the lattice constant.)  Dashed black lines indicate the phonon window $\mathcal{W}$, and the equilibrium retarded self-energy is shown as a black solid line.  Inset:  Semilog blowup of the region around $\omega=+\Omega$ showing the isosbectic
	point.
	}
	\label{fig:decay_vs_A}
\end{figure}
The changes in the scattering phase space due to the rearrangement of spectral weight by the pump
(as shown in Fig.~\ref{fig:cartoon}) imply that the measured decay rates depend on the pump fluence.
Thus,
we now explicitly consider the dependence of the effective electron-phonon interactions on the pump fluence or excitation density.  In particular, the analysis of Ref.~\cite{m_sentef_13} is repeated, and the decay rates 
are extracted from tr-ARPES spectra integrated over a cut along the $(11)$ momentum direction [as in Figs.~\ref{fig:fig1}(a) and (b)] for various pump fluences.  To be able to extract the decay rates from the spectra, sufficient signal is needed for an exponential fit.  
At the strong coupling considered above, the signal decays too rapidly, which is remedied by decreasing the coupling strength, and increasing both the driving frequency and temperature. 
Fig.~\ref{fig:decay_vs_A} shows the decay rates obtained just after the pump pulse, together with the equilibrium result.  The decay rates directly reflect the changes discussed in Fig.~\ref{fig:cartoon}; compared to equilibrium, the scattering rates increase inside $\mathcal{W}$ and decrease outside $\mathcal{W}$.  As the fluence increases, the rates deviate further from their values in equilibrium.  
Around the phonon frequency, there is a nearly isosbestic crossover point where the modifications in the
scattering rate change sign, which is shown in the inset of Fig.~\ref{fig:decay_vs_A}.

%\section{Figure 4}
\begin{figure}
	\includegraphics[width=\columnwidth]{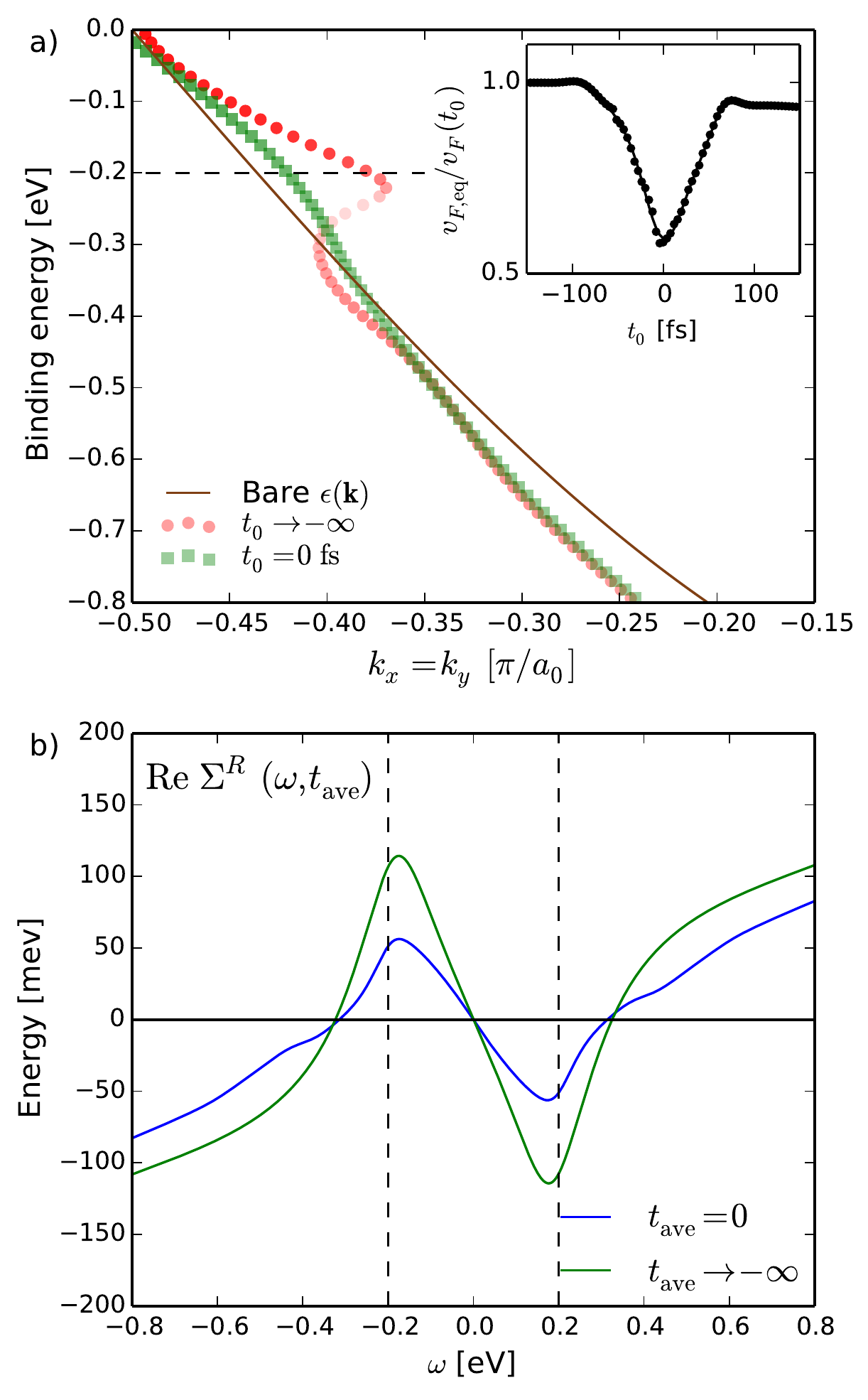}
	\caption{a) MDC peak positions from Fig.~\ref{fig:fig1}, together with the bare dispersion (solid line).  The inset shows the inverse Fermi velocity normalized by the equilibrium value (see text).  b) Real part of the Wigner self-energy $\Sigma^R(\omega,\tave)$ in equilibrium $(\tave\rightarrow -\infty)$  and at 0 time delay showing a decrease in peak height which leads to a straightening of the band in panel (a).  Dashed black lines indicate the phonon window $\mathcal{W}$.}
	\label{fig:fig4}
\end{figure}

Finally, we return to the weakening of the kink in the tr-ARPES spectra.
The dispersion is determined from Fig.~\ref{fig:fig1} by fitting the MDCs with a Lorentzian (as discussed above) and extracting the peak position.  Figure~\ref{fig:fig4}(a) shows the dispersion in equilibrium ($t_0\rightarrow -\infty$) and at $t_0=0$.  Clearly, the kink is much more pronounced in equilibrium then during the pump.  To get a measure of the kink as a function of time, the Fermi velocity is extracted by fitting the dispersion
near the Fermi level to a line.
The inset of Fig.~\ref{fig:fig4}(a) shows the inverse of the velocity at the Fermi level ($1/v_F(t_0)$), normalized to the equilibrium value ($1/v_{F,\mathrm{eq}}$).  The kink is suppressed strongly at zero time delay, when the field amplitude is largest. 
To explain the changes in the kink,
 we calculate the real part of the Wigner self-energy (defined above), which is plotted in Fig.~\ref{fig:fig4}(b).  The strength of the kink is related to the peaks in $\Sigma^R(\omega,\tave)$, which clearly decrease due to the pump.  
 
\section{Discussion}

Our previous studies\cite{a_kemper_13,m_sentef_13} have linked the decay rates and oscillation frequencies to the equilibrium self-energy in the limit of weak excitation, which is achieved either by going to long times or low pump fluences.  
At larger fluences, it is tempting to continue to discuss the pump-induced changes in the
effective electron-phonon interaction in terms of equilibrium physics.  In particular, the changes in the kink are strikingly similar to a situation where either the electron-phonon coupling strength is decreased, or the temperature is elevated.  One could speak in the first case about the decoupling of electrons from the phonons, in the second case about an elevated electron temperature.  As we have shown, neither situation is correct in discussing the pump-induced changes.  

% This belongs to below, but is here for formatting purposes
\begin{figure*}[htpb]	
	\includegraphics[clip=true,trim=0 0 0 0,width=\textwidth]{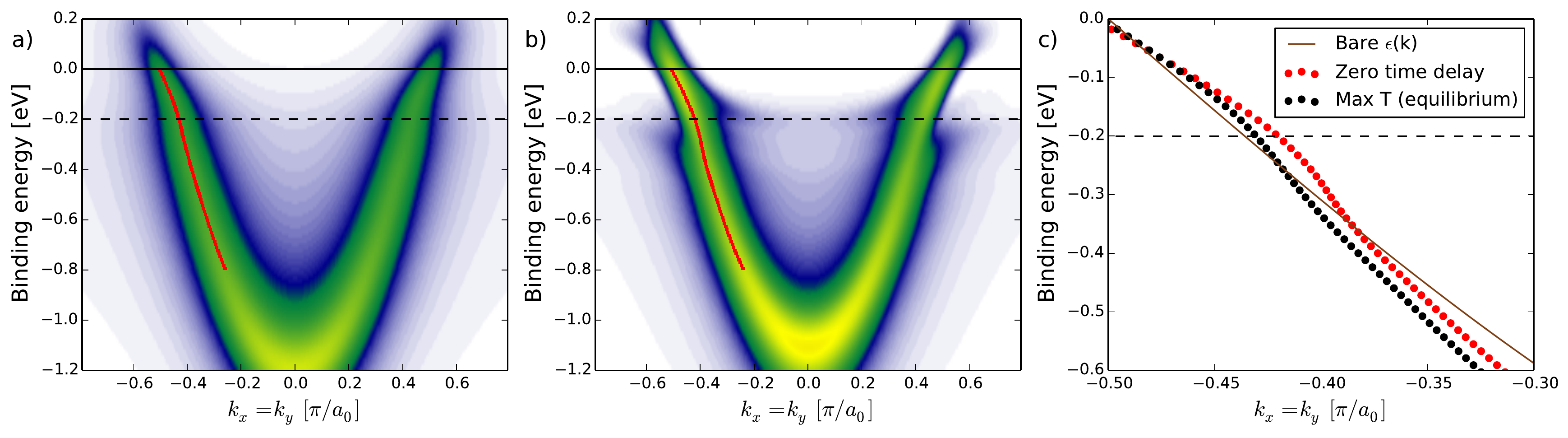}
	%\includesvg{figures/arpes_slices_mdc}
	\caption{a) tr-ARPES spectrum along the $(k_x=k_y)$ direction in equilibrium at $T=1500$ K.  b) Pumped tr-ARPES spectrum at zero time delay (reproduced from Fig.~1).  c) MDC centers from a) and b), together with the bare dispersion. 
	}
	\label{fig:arpes_T_comparison}
\end{figure*}

A similar decrease in the kink strength can be expected from an increase in the sample temperature; however, there are key differences in such a scenario. 
An increase in temperature would lead to a change in the sum rule, which depends on temperature through
the Bose function.  
As shown in Ref.~\onlinecite{m_sentef_13}, the photoemission spectra on the unoccupied side of the Fermi
level do not reflect
a heated Fermi-Dirac distribution.  Nevertheless, we extract the maximum temperature by fitting $\bar{I}(\omega,t_0)$ to a Fermi function close to the Fermi level, yielding a maximum observed temperature of $1500$ K.
 Figure~\ref{fig:arpes_T_comparison} shows the obtained high-temperature spectra, as well as the pumped spectrum from the main text at zero time delay.  At high temperatures, the self-energy is smeared out significantly, leading to the rather broad spectra [Fig.~\ref{fig:arpes_T_comparison}(a)].  On the other hand, the pumped spectra [Fig.~\ref{fig:arpes_T_comparison}(b)] show features reminiscent of the equilibrium strong coupling result \textemdash smaller line widths, as well as a remnant kink at the phonon frequency.  Figure~\ref{fig:arpes_T_comparison}(c) shows the MDC maxima extracted from the panels (a) and (b) of Fig.~1,
illustrating that the effective interaction is still quite different from a simple elevated temperature scenario, even
at the level of the dispersion.
 
We can further eliminate the quantum quench scenario because there is a frequency-dependent
reorganization of the effective interactions.
Instead of an overall decrease of the scattering rate, as one would expect if the electrons were decoupled from the phonons, we observe an increase within the phonon window $\mathcal{W}$.
If none of the physical parameters were changed, the sum rule would display a time dependence,
which is absent in our calculations.

The results presented here provide context for the analysis of other effects which are not included in the
model.  In the single band case, the interactions are determined by the spectral weight distribution
in the Green's function.  When multiple bands are present, the scattering between them must be taken
into account in the description of the interactions.  While the multiple bands can enter the interaction
in non-trivial ways, the same framework used in the analysis for a single band applies.

In equilibrium, the decrease of interactions at low energies would lead to a weakening or disappearance
of the emergent phenomena which depend on the interactions at those scales.  In non-equilibrium,
this is not yet known.  We have shown that the pump can cause a non-trivial
redistribution of the effective interactions to different energies.  This behavior can not be
captured by an elevated temperature or quantum quench picture.  Instead, the spectral weight
redistribution is critical to understanding the non-equilibrium response, and will have similar relevance
in pumping 
systems with emergent states.  It is again tempting to describe the effective interactions
within an equilibrium picture with modified parameters;  however, to capture the full physical description,
the non-equilibrium process should be considered within the framework presented here.

\appendix
\section{Details of the approach}
\label{sec:app_model}

In our calculations, the electron Green's functions are determined fully self-consistently at each time step.  We use the approximation of an infinite phonon bath whose properties are equal to the equilibrium noninteracting bath.  Mathematically, this means that the phonon propagator is the bare one, and is not updated in the calculation.  The renormalization of the phonon propagator is normally weak; additionally,
changes in the electronic spectral weight distribution only couple to the phonon propagator in second order,
and will not cause large changes.

To solve the equations of motion, we utilize  an expanding contour method described in Ref.~\onlinecite{a_stan_09}.
The equations of motion separate on the Keldysh contour into a set of equations for the various components of the Green's function, depending on the relative locations of the two times.  The contour-ordered
Green's function is
\begin{align}
G_\kk^\mathcal{C}(t,t') = -i \langle \mathcal {T_C}\, c_\kk(t) c_\kk^\dagger(t') \rangle
\end{align}
where $t$ and $t'$ lie on the Keldysh contour, $\mathcal{T_C}$ is the contour time-ordering
function.
For completeness, the equations of motion are (letting the contour start at $\tmin$),
\begin{widetext}
\begin{subequations}
\begin{align}
\big[ -\partial_\tau - \epsilon(\kk(\tmin))\big] G_\kk^M(\tau) =& \delta(\tau) + \int_0^\beta d\bar \tau\  \Sigma^M(\tau-\bar\tau) G_\kk^M(\bar\tau), \\
\big[ i\partial_t - \epsilon(\kk(t))\big] G_\kk^\rceil(t,-i\tau) =& \int_{\tmin}^t d\bar t\  \Sigma^R(t,\bar t) G_\kk^\rceil(\bar t, -i\tau) 
+ \int_0^\beta d\bar\tau\ \Sigma^\rceil(t,-i\bar\tau) G_\kk^M(\bar\tau -\tau),\\
\big[ i\partial_t - \epsilon(\kk(t))\big] G_\kk^\gtrless(t,t') =& \int_{\tmin}^t d\bar t\  \Sigma^R(t,\bar t) G_\kk^\gtrless(\bar t, t') 
+\int_{\tmin}^{t'} d\bar t\  \Sigma^\gtrless(t, \bar t) G_\kk^A(\bar t, t') %\nonumber 
 - i \int_0^\beta d\bar\tau\ \Sigma^\rceil(t,-i\bar\tau) G_\kk^\lceil(-i\bar\tau,t').
 \label{eq:Gless}
\end{align}
\label{eq:eoms}
\end{subequations}
\end{widetext}
The superscripts $M,\rceil,\lceil,R,A\gtrless$ indicate the Matsubara, mixed real-imaginary, mixed imaginary-real, retarded, advanced, and greater/lesser components, respectively.  Similar equations are obtained by using $t'$ instead of $t$.  The various components can be transformed or combined into others by the usual relations\cite{mahan}.

\begin{figure}[htpb]	
	\includegraphics[clip=true,trim=0 0 0 0,width=0.8\columnwidth]{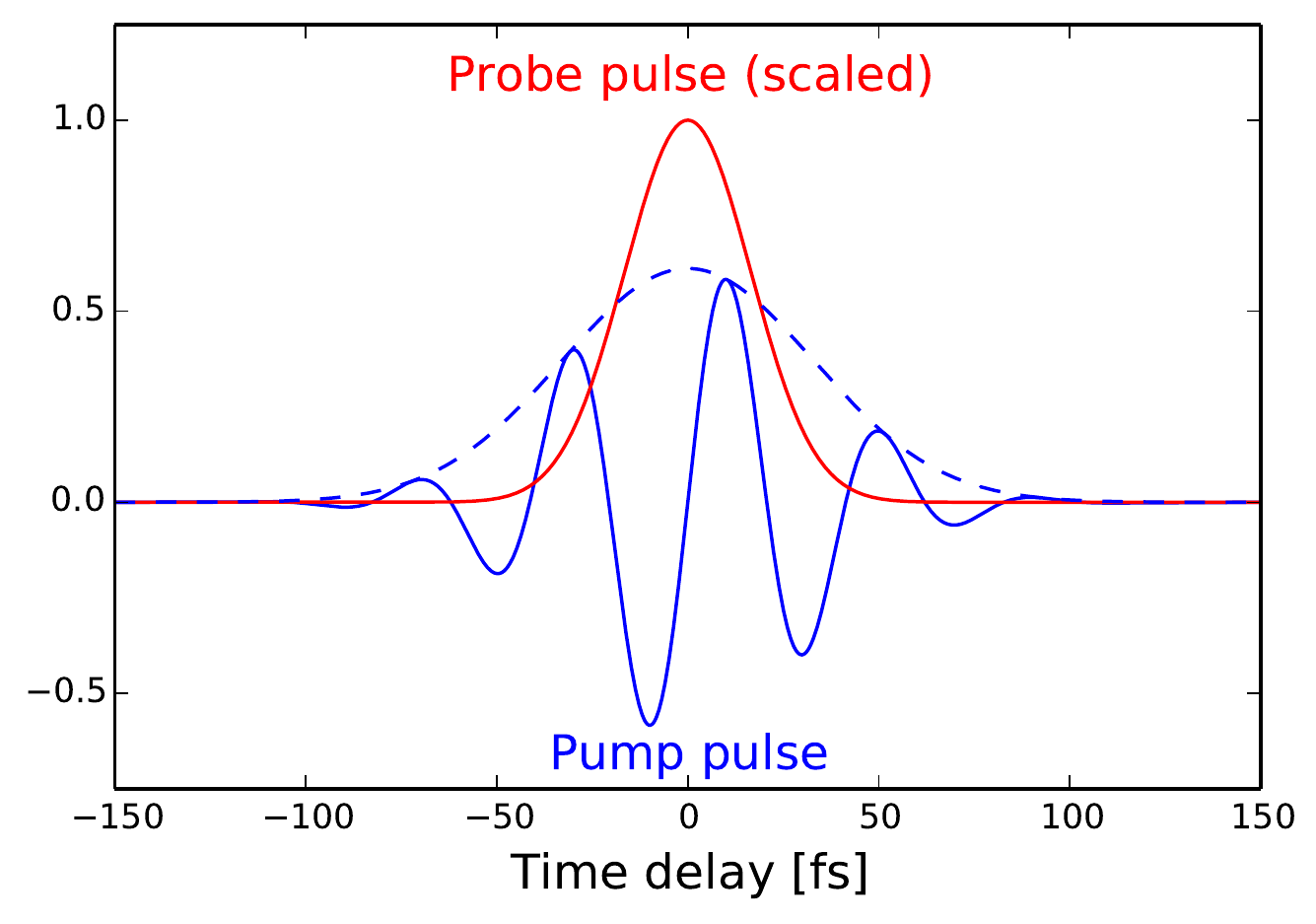}
	\caption{Pump and probe fields used in the calculation of the tr-ARPES spectra.}
	\label{fig:field}
\end{figure}

Once the Green's functions are obtained, time-resolved ARPES (tr-ARPES) spectra can be computed.  For a probe pulse of width $\sigma_p$, the tr-ARPES intensity at time $t_0$ is \cite{j_freericks_09}

\begin{align}
I(\kk,\omega,t_0) =\mathrm{Im} \int dt\, dt'\, 
p(t,t',t_0)
e^{i\omega(t-t')} G_{\tilde\kk}^<(t,t') 
\end{align}
where $p(t,t',t_0)$ is a two-dimensional normalized Gaussian with a width $\sigma_p$ centered at $(t,t')=(t_0,t_0)$.  The field-induced shift in $\kk$  has to be corrected via a gauge shift in the momentum argument of $G^<_\kk$ with \cite{bertoncini_91,v_turkowski_book}
\begin{align}
\tilde\kk = \kk + \frac{1}{t-t'} \int_{t'}^{t} d\bar t\, \A(\bar t).
\end{align}
To determine the tr-ARPES spectral weight, we utilize a probe with a Gaussian envelope whose width $\sigma_p=16.45$ fs (Fig.~\ref{fig:field}).

From Eq.~\ref{eq:Gless} we can identify the components contributing to the return to equilibrium of the tr-ARPES spectra.  Drawing an analogy to a simple first order differential equation, we observe that the retarded self-energy plays the role of a time constant, and the remaining integrals appear as driving terms.  Thus, while the retarded self-energy governs a substantial portion of the temporal dynamics, it is not the full story and the other pieces must be taken into account.  This implies that the decay rate is not 100\% determined by the retarded self-energy outside of the weak excitation limit.

The Wigner self-energy, which is used above, is defined as follows.  First, a transformation is made from $(t,t')$ coordinates to $(\trel,\tave)$ coordinates via $\tave = [t+t']/2$ and $\trel=t-t'$.  A Fourier transform $\mathcal{F}$ on $\trel$ gives $\Sigma(\omega,\tave) = \mathcal{F}\left[\Sigma(\trel,\tave)\right]$.

\begin{widetext}
\ \\
\section{Sum rule for the self-energy}
\label{sec:app_sumrule}
\begin{figure}[htpb]	
	\includegraphics[clip=true,trim=0 0 0 0,width=0.4\textwidth]{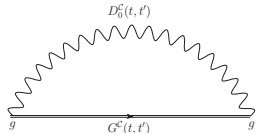}
	\caption{Feynman diagram for the contour self-energy.}
	\label{fig:feynman}
\end{figure}
For the Holstein model with an unrenormalized phonon, the contour self-energy (see Fig.~\ref{fig:feynman}) is given by
\begin{align}
\Sigma^\mathcal{C}(t,t') = \frac{i g^2}{ N_\kk} \sum_\kk G_\kk^\mathcal{C}(t,t')  D^\mathcal{C}_0(t,t'),
\label{eq:sigma_contour}
\end{align}
where 
$N_\kk$ is the number of momenta. The zeroth frequency moment for the self-energy (in frequency) is related to the diagonal part of $\Sigma^R(t,t')$.   We use the bare contour phonon propagator\cite{mahan}
\begin{align}
D^\mathcal{C}_0(t,t') =& -i \big[ n_B(\Omega/T) + 1 - \theta_\mathcal{C}(t,t') \big] e^{i\Omega(t-t')} \nonumber \\
& -i \big[ n_B(\Omega/T) + \theta_\mathcal{C}(t,t') \big] e^{-i\Omega(t-t')},
\end{align}
where $n_B(x)$ is the Bose function $n_B(x) = \left[e^x-1\right]^{-1}$, and $\theta_\mathcal{C}(t,t')$ is the contour Heaviside function.  Using Eq.~\ref{eq:sigma_contour}, we extract the retarded part of $\Sigma^\mathcal{C}$\cite{r_vanleeuwen_05} and set $t'=t$,

\begin{align}
%\Sigma^R(t,t'=t) = \sum_\kk \left[G_\kk^>(t,t) - G_\kk^<(t,t)\right] D_0^<(t,t) \nonumber \\+ G^<_\kk(t,t) D^A(t,t)
\Sigma^R(t,t) &= \frac{i g^2}{ N_\kk}  \sum_\kk \bigg[ G_\kk^R(t,t) D_0^>(t,t)+ G^<_\kk(t,t) D_0^R(t,t) \bigg],\\
&=  \frac{i g^2}{ N_\kk} \sum_\kk  \bigg[ (-i) \cdot \left[ -i(2n_B(\Omega/T)+1)  \right] + G^<_\kk(t,t) \cdot 0 \bigg], \\
&= -ig^2\big[2n_B(\Omega/T)+1\big].
\end{align}
\end{widetext}
Thus, as long as the phonon occupation and the bare vertex are unchanged, the sum rule for the self-energy is preserved.  In our simulations, this sum rule is found to hold numerically, indicating that the total electron-phonon coupling strength remains fixed.

%\section{Comparison to high temperature}
%As the spectra are not well represented by Fermi functions over a wide energy range, one immediately sees that the resulting temperature would be an uncertain quantity at best \cite{m_sentef_13}.  Nevertheless, we extract the maximum temperature by fitting $\bar{I}(\omega,t_0)$ to a Fermi function close to the Fermi level, yielding a maximum observed temperature of $1500$ K.  

\begin{acknowledgments}
We would like to thank P. Kirchmann, J. Sobota and S. Yang for helpful discussions.  This work was supported by the Department of Energy, Office of Basic Energy Sciences, Division of Materials Sciences and Engineering (DMSE) under Contract Nos. DE-AC02-76SF00515 (Stanford/SIMES), DE-FG02-08ER46542 (Georgetown), and DE-SC0007091 (for the collaboration). Computational resources were provided by the National Energy Research Scientific Computing Center supported by the Department of Energy, Office of Science, under Contract No. DE- AC02-05CH11231. J.K.F. was supported by the McDevitt bequest at Georgetown.
\end{acknowledgments}

\bibliography{master,michael}

\end{document}